\documentclass[a4paper,11pt]{article}
\usepackage{pos}

\newcommand{\mua}{\mu}
\newcommand{\ma}{m}

\usepackage[normalem]{ulem}  % \sout{old text} for strikeout
\title{Perturbative predictions for color superconductivity on the lattice}
\author*[a,b]{Takeru Yokota}
\author[c]{Yuhma Asano}
\author[d]{Yuta Ito}
\author[e,f]{Hideo Matsufuru}
\author[g]{Yusuke Namekawa}
\author[e,f]{Jun Nishimura}
\author[h]{Asato Tsuchiya}
\author[i]{Shoichiro Tsutsui}

\affiliation[a]{
Interdisciplinary Theoretical and Mathematical Sciences Program (iTHEMS), RIKEN, 
Wako, Saitama 351-0198, Japan}
\affiliation[b]{Institute for Solid State Physics, The University of Tokyo,
Kashiwa, Chiba 277-8581, Japan}
\affiliation[c]{Faculty of Pure and Applied Sciences,
University of Tsukuba, 
1-1-1 Tennodai, Tsukuba, Ibaraki 305-8577 Japan}
\affiliation[d]{National Institute of Technology, Tokuyama College, 
Gakuendai, Shunan, Yamaguchi 745-8585, Japan}
\affiliation[e]{High Energy Accelerator Research Organization (KEK), 
  1-1 Oho, Tsukuba, Ibaraki 305-0801, Japan}
\affiliation[f]{School of High Energy Accelerator Science, Graduate University for Advanced Studies (SOKENDAI), 
1-1 Oho, Tsukuba, Ibaraki 305-0801, Japan}
\affiliation[g]{Department of Physics, Kyoto University, Kyoto 606-8502, Japan}
\affiliation[h]{Department of Physics, Shizuoka University, 
836 Ohya, Suruga-ku, Shizuoka 422-8529, Japan}
\affiliation[i]{Quantum Hadron Physics Laboratory, RIKEN Nishina Center, Wako, Saitama 351-0198, Japan}

\emailAdd{takeru.yokota@riken.jp}
\abstract{
  We develop a new method to investigate color superconductivity (CSC)
  on the lattice based on the Thouless criterion,
  which amounts to solving the linearized gap equation
  without imposing any ansatz on the structure of the Cooper pairs.
  We perform explicit calculations at the one-loop level
  with the staggered fermions on a $8^3 \times 128$ lattice
  and the Wilson fermions on a $4^3 \times 128$ lattice,
  which enables us to obtain the critical $\beta(=6/g^2)$ 
  as a function of the quark chemical potential $\mua$,
  below which the CSC phase is expected to appear.
  The obtained critical $\beta$ has sharp peaks at the values of $\mua$ corresponding to
  the discretized energy levels of quarks similarly to what was observed
  in previous studies on simplified effective models.
  From the solution to the linearized gap equation,
  one can read off the flavor and spatial structures of the Cooper pairs at the critical $\beta$.
  In the case of massless staggered fermion, in particular,
  we find that the chiral $\mathrm{U}(1)$ symmetry of the staggered fermions
  is spontaneously broken by the condensation of the Cooper pairs.
}

\FullConference{%
 The 38th International Symposium on Lattice Field Theory, LATTICE2021
  26th-30th July, 2021
  Zoom/Gather@Massachusetts Institute of Technology
}

\note{RIKEN-iTHEMS-Report-21, RIKEN-QHP-505, KEK-TH-2360, KUNS-2897}

\begin{document}
\maketitle

\section{Introduction}
The QCD phase diagram
%quantum chromodynamics (QCD)
is expected to have a rich structure, whose 
elucidation is one of the biggest goals in high energy physics.
%For that, lattice Monte Carlo simulation should play an important role
%as a powerful first-principles method for QCD.
While the first-principles studies of the dense QCD matter is known to be extremely difficult
due to the notorious sign problem,
%a long-standing problem
%because lattice Monte Carlo simulations fail in dense systems 
there are various methods that have been developed in recent years
to circumvent this problem.
In particular, the complex Langevin method \cite{par83,kla84}
has been applied to lattice QCD at finite density with promising results;
see Ref.~\cite{Berger:2019odf} and references therein.
%\cite{Sexty:2013ica,Aarts:2014bwa,Fodor:2015doa,Sinclair:2015kva,Sinclair:2016nbg,Sinclair:2017zhn,Sinclair:2018rbk,Nagata:2018mkb,Ito:2018jpo,Tsutsui:2018jva,doi:10.7566/JPSCP.26.024012,Kogut:2019qmi,Sexty:2019vqx,Sinclair:2019ysx,Tsutsui:2019suq,Scherzer:2020kiu,Ito:2020mys}

%,
%which include the complex Langevin method \cite{par83,kla84}, Lefschetz thimble method \com{[]},
%tensor renormalization group \com{[]},
%and path optimization \com{[]}.
%As the application of the complex Langevin method to lattice QCD \com{[]}
%is realized for instance,
%these developments give a hope
%for the understanding of the QCD phase diagram
%from first principles.

In fact, the QCD phase diagram 
can be investigated in the high density regime
by perturbation theory
thanks to the asymptotic freedom.
In particular, it is expected that
the color superconductivity (CSC) \cite{bar77,fra80,bai81,alf98,raj00}
occurs in the cold dense region
considering that the color-anti-triplet channel 
of the potential induced by one-gluon exchange is attractive.
Qualitative properties of the CSC
such as the scaling of the gap function with respect to the coupling constant
have been discussed based on perturbative QCD \cite{alf08}.

Perturbative studies are expected to be useful
not only in exploring the QCD phase diagram at high density but also
in providing predictions for
%the results obtained by
first-principles calculations in the cold dense region,
which seem to be quite promising \cite{Ito:2020mys}.
While quantitative predictions on the parameter region in which the CSC occurs
were discussed in simplified effective models 
such as the Nambu--Jona-Lasinio-like model \cite{alf98,raj00},
there have been no such works in QCD.
%still remain as difficult problems,
One of the
%difficulties
reasons for this
%of such quantitative studies of the CSC
is that one has to solve a nonlinear functional equation known as the gap equation,
%, which plays a central role in perturbative studies of the CSC,
%for the gap function,
which is extremely difficult, in particular,
without imposing some ansatz on the gap function.
%almost impossible
%unless one knows in advance a natural ansatz to be imposed on the gap function.

%% However, this difficulty is relaxed on finite lattice,
%% since the gap equation is reduced to
%% finite number of coupled equations.
%% Moreover, lattice setup has an advantage that
%% lattice regularization does not break the gauge symmetry.

In this paper, we investigate the CSC
by solving the gap equation on the lattice
%study CSC numerically
without imposing any ansatz on the structure of the gap function describing
the condensation of the Cooper pairs.
This is possible since the gap equation is reduced on the lattice
to a finite number of coupled equations.
%Also the lattice regularization has an advantage that it respects the gauge symmetry
%unlike the previously adopted momentum cutoff regularization.
Further simplification of the gap equation is achieved by
focusing on the critical point so that the gap equation can be linearized.
%is reduced to a linear equation for the gap function.
The condition for the linearized gap equation to have a nontrivial solution
is well known in the condensed matter physics
as the Thouless criterion, but to our knowledge, this is the first time that it was
applied to the CSC in QCD.
%Particularly, we focus on critical points, where the gap equation is reduced to a linear equation.

Thus, our method enables us
%to calculate the critical $\beta=6/g^2$
%corresponding to the critical point and
to provide a quantitative prediction on the parameter region of the CSC by calculating
the critical coupling $\beta=6/g^2$, which we denote as $\beta_{\rm c}$ in what follows,
as a function of the quark chemical potential for the staggered and Wilson fermions.
The results for $\beta_{\rm c}$
show peak structure as a function of the chemical potential,
which is due to the discretized energy levels of quarks in a finite volume.
We also investigate the structure of the Cooper pairs
at $\beta_{\rm c}$
from the solution to the linearized gap equation
in the case of massless staggered fermions.
%eigenvector corresponding to 
From the results for scalar and pseudo-scalar condensates,
we find that the chiral $\mathrm{U}(1)$ symmetry is broken
spontaneously.
%by the condensates.
The spatial structure of the Cooper pairs, on the other hand,
exhibits behaviors consistent with the BCS theory of superconductivity.

The rest of this paper is organized as follows.
In Section~\ref{sec:form}, we present our general formalism for the CSC.
In Section~\ref{sec:bc_stag},
we show our numerical results for $\beta_{\rm c}$ in the case of staggered fermions,
followed by Section~\ref{sec:cooper_stag},
where we investigate the structure of the Cooper pairs at the critical point
in the massless case.
%case of massless staggered fermions.
In particular, we identify the flavor and spatial structures of the Cooper pairs
and discuss the spontaneous breaking of chiral $\mathrm{U}(1)$ symmetry.
In Section \ref{sec:bc_wil}, we present our results for $\beta_{\rm c}$
in the case of Wilson fermions.
Section \ref{sec:conc} is devoted to a conclusion.

\section{General formalism for CSC on the lattice\label{sec:form}}

Since we are going to investigate CSC on the lattice
for staggered and Wilson fermions,
here we describe a general formalism, which is applicable to both cases.
We introduce a fermion field $\psi^{a}_{\rho}(n)$,
%and interacts with gluons.
where $n$ is the label of sites on the lattice,
and $a$ is the color index.
The index $\rho$ represents the internal degrees of freedom
other than color such as the flavor index and the spinor index collectively.
We also introduce the conjugate field $\overline{\psi}^{a}_{\rho}(n)$.

In order to investigate the condensate of Cooper pairs,
we use the Nambu-Gorkov formalism,
in which one introduces
the Nambu basis $\Psi_\rho^a(n)=(\psi_\rho^a(n),\overline{\psi}_\rho^a(n))^{\rm t}$ \cite{nam60,bar77,bai81}.
Assuming that the lattice translational symmetry is not spontaneously broken,
it is convenient to work with the momentum representation
since two-point correlation functions are diagonalized
with respect to momenta owing to the momentum conservation.
The propagator for the Nambu basis $\tilde{\bf S}^{aa'}_{\rho\rho'}(p)$
in the momentum representation
%with $p$ being the momentum
can be represented by a $2\times2$ matrix, which satisfies the Dyson equation
%given as
\begin{align}
	\label{eq:dyson}
	\tilde{\bf S}^{-1,aa'}_{\rho\rho'}(p)
	= \tilde{\bf D}_{\rho\rho'}^{aa'}(p) + \tilde{\bf \Sigma}_{\rho\rho'}^{aa'}(p) \ ,
\end{align}
where $\tilde{\bf D}_{\rho\rho'}^{aa'}(p)$
is the inverse free propagator defined by 
$\tilde{\bf D}_{\rho\rho'}^{aa'}(p)=
{\rm diag}(\tilde{D}_{\rho\rho'}^{aa'}(p),
-\tilde{D}_{\rho'\rho}^{aa'}(-p))$
with $\tilde{D}_{\rho\rho'}^{aa'}(p)$ being
the inverse free propagator for $\psi^{a}_{\rho}(n)$
and $\tilde{\bf \Sigma}_{\rho\rho'}^{aa'}(p)$
being the self-energy defined as
\begin{align}
	\tilde{\bf \Sigma}_{\rho\rho'}^{aa'}(p)
	=
	\begin{pmatrix}
		\tilde{\Sigma}_{11,\rho\rho'}^{aa'}(p)
		&
		\tilde{\Sigma}_{12,\rho\rho'}^{aa'}(p)
		\\
		\tilde{\Sigma}_{21,\rho\rho'}^{aa'}(p)
		&
		\tilde{\Sigma}_{22,\rho\rho'}^{aa'}(p)
	\end{pmatrix} \ .
\end{align}
A nonvanishing off-diagonal part of the self-energy
$\tilde{\Sigma}_{12(21),\rho\rho'}^{aa'}(p)$
corresponds to the appearance of pair condensate,
which is crucial for the CSC,
%since they indicate the appearance of pair condensate.
while the diagonal part is considered as higher-order contributions in our calculation,
which is neglected in what follows.
%Hereafter, we ignore which 

\begin{figure}[!t]
  \begin{center}
	\begin{alignat*}{3}
	\mathrm{(a)}
	\qquad
	&
	\tilde{\Sigma}_{12,\rho\rho'}^{aa'}(p)
	&=
	\parbox[c]{12em}{\includegraphics[width=12em]{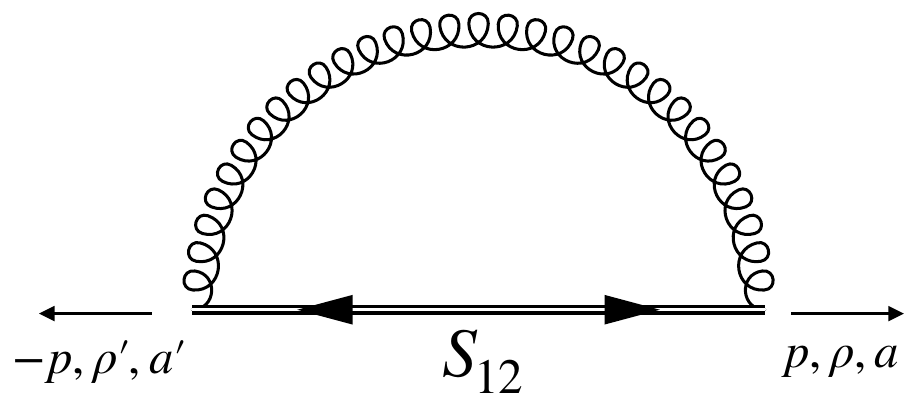}}
	\\
	\mathrm{(b)}
	\qquad
	&
	\frac{1}{\beta} \mathcal{M}_{(p\rho\rho')(q\sigma\sigma')}
	&=
	\parbox[c]{12em}{\includegraphics[width=12em]{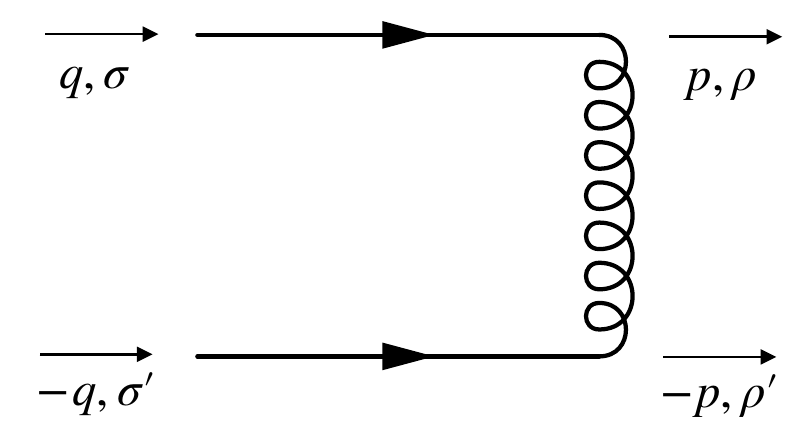}}
	\end{alignat*}
    \caption{Diagrammatic representation of (a) the gap equation for 
    $\tilde{\Sigma}_{12,\rho\rho'}^{aa'}(p)$
    and (b) $\mathcal{M}_{(p\rho\rho')(q\sigma\sigma')}$.
    The coiled and solid lines stand for the gluon propagator and
    the free fermion propagator $\tilde{D}^{-1,aa'}_{\rho\rho'}(p)$, respectively.}
    \label{fig:sm_diag}
  \end{center}
\end{figure}
At the one-loop level, the self-consistency equation
for $\tilde{\Sigma}_{12,\rho\rho'}^{aa'}(p)$ or the gap equation
%i.e., the gap equation
is given by Fig.~\ref{fig:sm_diag} (a).
%% In both cases of the staggered and Wilson fermions,
%% the Feynman rules to convert the diagrams into equations
%% are immediately obtained from their actions.
The gap equation for the other off-diagonal part $\tilde{\Sigma}_{21,\rho\rho'}^{aa'}(p)$
need not be considered since it gives the same results as the one
for 
%can be obtained in the same manner, we focus on 
$\tilde{\Sigma}_{12,\rho\rho'}^{aa'}(p)$.
%since they give the same results.
While one can obtain $\tilde{\Sigma}_{12,\rho\rho'}^{aa'}(p)$ in principle
by solving the gap equation together with the Dyson equation \eqref{eq:dyson}, 
it does not seem to be feasible to do so in practice
%infeasible to find nonzero solutions of the gap equation
without having a natural ansatz to impose
on $\tilde{\Sigma}_{12}(p)$ for arbitrary parameters.
Instead, we
%consider the so-called Thouless criterion, which amounts to focusing
focus on the critical point at which $\tilde{\Sigma}_{12}(p)\to 0$
assuming a second-order phase transition.
This reduces the gap equation to a linear equation,
which can be solved without imposing any ansatz on $\tilde{\Sigma}_{12}(p)$.
%In this case, because of $\tilde{\Sigma}_{12}(p)\to 0$,
%the gap equation is reduced to a linear equation.
Here we consider the color anti-symmetric component 
$\sum_{ab}\epsilon_{abc}\tilde{\Sigma}_{12,\rho\rho'}^{ab}(p)$,
for which the interaction is attractive and hence the Cooper instability is expected.
Since a different choice of $c$ simply leads to the same equation,
we choose $c=3$ without loss of generality and define
$\tilde{\Sigma}^{(-)}_{12(p\rho\rho')}
= \sum_{ab}\epsilon_{ab3}\tilde{\Sigma}_{12,\rho\rho'}^{ab}(p) $,
which satisfies the linearized gap equation
\begin{align}
	\label{eq:gap_linear}
	\sum_{q\sigma\sigma'}
	\mathcal{M}_{(p\rho\rho')(q\sigma\sigma')}
	\tilde{\Sigma}^{(-)}_{12(q\sigma\sigma')}
	=
	\beta \, 
	\tilde{\Sigma}^{(-)}_{12(p\rho\rho')} \ .
\end{align}
We have introduced $\beta=2N_{\rm c}/g^2$, where
$g$ is the gauge coupling constant with $N_{\rm c}=3$
in the case at hand.
%is the color degree of freedom for fermions.
The matrix $\mathcal{M}_{(p\rho\rho')(q\sigma\sigma')}$
is independent of $\beta$,
and it is given diagrammatically in Fig.~\ref{fig:sm_diag} (b).
From Eq.~\eqref{eq:gap_linear},
%we find that $\beta$ is given as an eigenvalue of
%$\mathcal{M}$.
%In particular, we have
one finds that the largest eigenvalue of $\mathcal{M}$
gives the critical point
\begin{align}
	\beta_{\rm c}=\lambda_{\rm max}[\mathcal{M}] \ ,
\end{align}
which determines the boundary of
%i.e., the interface between
the normal and superconducting phases.
%% where $\lambda_{\rm max}[\mathcal{M}]$
%% is the largest eigenvalue of $\mathcal{M}$
%% and $\beta_{\rm c}$
%% corresponds to the critical point,
%% i.e., the interface 
%% between the normal and superconducting phases.
The eigenvector $\tilde{\Sigma}^{(-)}_{12(p\rho\rho')}$
corresponding to the largest eigenvalue, on the other hand,
%$\lambda_{\rm max}[\mathcal{M}]$. 
tells us the structure of the Cooper pairs at the critical point.
Such a condition for the critical point is equivalent to
the condition for the divergence of the T-matrix,
which is
%forms the basis of the well-known
well known as the Thouless criterion \cite{tho60} in condensed matter physics.
Since the size of the matrix $\mathcal{M}$ is finite for a finite lattice,
%in the present case of a finite lattice, 
%is finite, which allows us to
we can calculate the largest eigenvalue $\lambda_{\rm max}[\mathcal{M}]$
and the corresponding eigenvector numerically by the standard power iteration method.
%for our numerical analysis.

%and numerical methods such as 
%the power iteration are applicable to calculate $\lambda_{\rm max}[\mathcal{M}]$.

\section{Critical coupling for the staggered fermions\label{sec:bc_stag}}
\begin{figure}[!t]
  \begin{center}
  	\includegraphics[width=0.5\columnwidth]{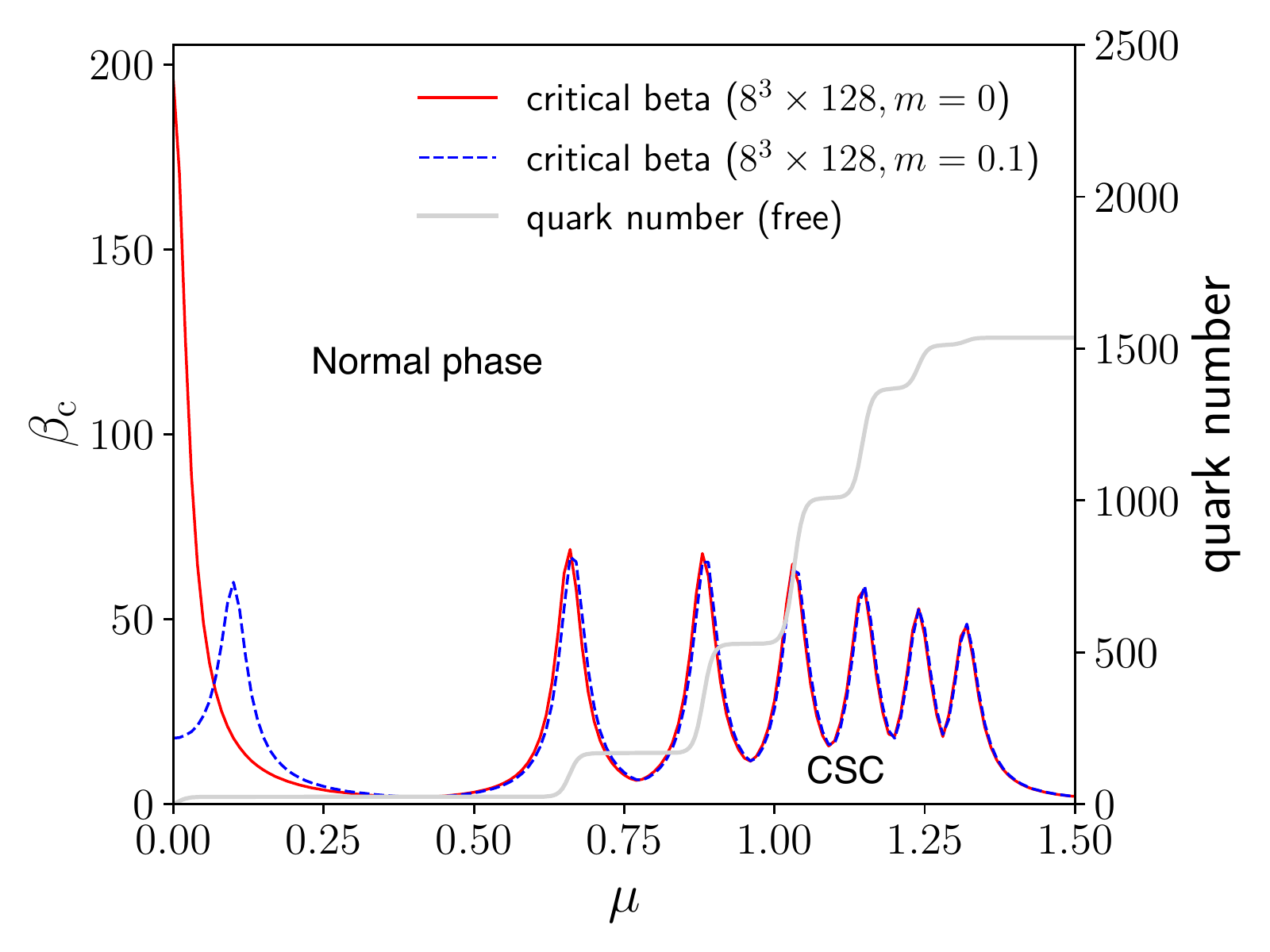}
    \caption{The critical coupling $\beta_{\rm c}$ is plotted as a function of $\mua$
      on an $8^3\times 128$ lattice using the staggered fermions
      with $\ma=0$ (red solid line) and $\ma=0.1$ (blue dashed line).
    The regions above and below these lines correspond to
    the normal and color-superconducting phases, respectively.
    The gray solid line represents
    the quark number for $\ma=0$ in the free case.}
        \label{fig:l8t128}
  \end{center}
\end{figure}
Let us first consider staggered fermions
and present our numerical results for $\beta_{\rm c}$
as a function of the quark chemical potential $\mua$
defined in lattice units.
%\sout{The power iteration method is employed for our calculation.}
Note that our calculation is valid at weak coupling,
which implies that the results are reliable if $\beta_{\rm c}\gg 1$.
We find that this is possible
%for sufficiently low temperature
%achieved with sufficiently low temperature realized by
%corresponding to a small
when the aspect ratio $L_{\rm s}/L_{\rm t}$ is sufficiently small,
where $L_{\rm s}$ and $L_{\rm t}$ represent the spatial and temporal extents of the lattice,
respectively.

Figure \ref{fig:l8t128} shows the result for an $8^3\times 128$ lattice
and the quark mass $\ma=0$ and $0.1$ in lattice units.
We also plot the number of quarks $N_{\rm q}$ in the same figure
for $\ma=0$ in the free case.
One can see that the peaks of $\beta_{\rm c}$ appear at the values of
$\mua$ for which $N_{\rm q}$ jumps from one plateau to another.
This can be understood theoretically as follows.
Let us note first that
the energy levels of free quarks
$E({\bf p})=\sinh^{-1}\sqrt{\sum_{i=1}^{3}\sin^{2} p_{i}+\ma^2}$
are discretized since the momentum ${\bf p}$ is discretized in a finite volume.
When the chemical potential $\mua$ goes beyond an energy level $E({\bf p})$,
$N_{\rm q}$ jumps because the number of momentum modes below the Fermi sphere increases.
Since there are momentum modes near the Fermi surface in this situation,  
the Cooper pairs are easy to form,
and hence the critical $\beta_{\rm c}$ has a peak at the same $\mua$.
%positions of are given by the energy levels of quarks.
This understanding is supported also by
the result for $\ma=0.1$ in Fig.~\ref{fig:l8t128},
where we observe that the peaks shift in accord with the shift of the energy levels.
%showing the shift of the first peak,
%which is consistent with the fact that $\ma$ shifts the energy levels.
Thus we find that the CSC region is suppressed when the chemical potential is not
close to any of the energy levels of quarks,
as has been also observed in the Nambu--Jona-Lasinio-like model \cite{amo02}.
The peak at $\mua \sim 0$ is considered to be a finite-size artifact
since the Cooper pairs in this case are formed
by quarks and anti-quarks with ${\bf p}={\bf 0}$,
whose density actually vanishes in the infinite volume limit.

\section{Structure of the Cooper pairs for the staggered fermions \label{sec:cooper_stag}}

As we mentioned in Section~\ref{sec:form},
the structure of the Cooper pairs
can be read off from the eigenvector $\tilde{\Sigma}_{12(p\rho\rho')}^{(-)}$
corresponding to $\lambda_{\rm max}[\mathcal{M}]$.
For that,
we rewrite $\tilde{\Sigma}_{12(p\rho\rho')}^{(-)}$ in Eq.~\eqref{eq:dyson}
into the anomalous correlation function of
the four-flavor Dirac fermion fields
in the coordinate space
using the relation between the staggered and Dirac fermions \cite{rot12}
and performing the Fourier transformation.
While our calculation is applicable to 
the Cooper pairs in any irreducible representations
of the Lorentz group,
let us focus, for simplicity, on the scalar (s) and pseudo scalar (ps) condensates
given as
\begin{align}
	\label{eq:ksps}
	K^{fg}_{\rm s(ps)}(n)
	=&
	\sum_{ab}
	\epsilon_{ab3}
	\left\langle
	{}^{\rm t}
	\psi^{a,f}(n)
	\mathcal{O}_{\rm s(ps)}
	\psi^{b,g}(0)
	\right\rangle \ .
\end{align}
Here $\psi^{a,f}(n)$ represents the Dirac fermion field on the coarse lattice
at a site $N$ with the color index $a$ and the flavor index $f$.
We have also introduced 
$\mathcal{O}_{\rm s}=\gamma_5 C$ and $\mathcal{O}_{\rm ps}=C$,
where $\gamma_5=\gamma_{1}\gamma_{2}\gamma_{3}\gamma_{4}$
and the charge conjugation operator $C=\gamma_2\gamma_4$
are defined in terms of the Euclidian Dirac gamma matrices $\gamma_\mu$.
Note that the overall normalization of $K^{fg}_{\rm s(ps)}(n)$
is irrelevant
%cannot be determined
since we are solving the linearized gap equation.

%\begin{figure}[!t]
%  \begin{center}
%  	\includegraphics[width=0.7\columnwidth]{fig_adep_mq0}
%    \caption{Flavor dependences of 
%    (a) scalar condensate $K_{\rm s}^{fg}(0)$ 
%    and (b) pseudo scalar condensate $K_{\rm ps}^{fg}(0)$
%    obtained from the eigenvector in the case of $\ma=0$.
%    All the components in Figs.~(a) and (b) are normalized by
%    $K_{\rm s}^{12}(0)$ and $K_{\rm ps}^{14}(0)$, respectively.
%    The color of the heat map corresponds to the absolute value
%    of each component.}
%    \label{fig:adep_mq0}
%  \end{center}
%\end{figure}
%
%\begin{figure}[!t]
%  \begin{center}
%  	\includegraphics[width=0.7\columnwidth]{fig_adep_mq0}
%    \caption{Absolute values
%    of the coefficients
%    when $K_{\rm s(ps)}^{fg}(0)$
%    is decomposed 
%    as
%    $
%K_{\rm s(ps)}^{fg}(0)
%    =
%    a_1 t_1^{fg}
%    +
%    a_3 t_3^{fg}
%    +
%    a_{13} \omega_{13}^{fg}
%    +
%    a_{24} \omega_{24}^{fg}
%    +
%    a_{25} (t_2t_5)^{fg}
%    +
%    a_{45} (t_4t_5)^{fg}
%    $.
%    Figures (a) and (b)
%    are the cases of
%    the scalar and pseudo-scalar
%    condensates, respectively.
%%    Flavor dependences of 
%%    (a) scalar condensate $K_{\rm s}^{fg}(0)$ 
%%    and (b) pseudo scalar condensate $K_{\rm ps}^{fg}(0)$
%%    obtained from the eigenvector in the case of $\ma=0$.
%%    All the components in Figs.~(a) and (b) are normalized by
%%    $K_{\rm s}^{12}(0)$ and $K_{\rm ps}^{14}(0)$, respectively.
%%    The color of the heat map corresponds to the absolute value
%%    of each component.
%}
%    \label{fig:adep_mq0}
%  \end{center}
%\end{figure}
Let us recall here that we are considering the case in which
the pair condensate is color-antisymmetric and
it is either a Lorentz scalar or a pseudo-scalar.
This forces $K_{\rm s(ps)}^{fg}(0)$ to be antisymmetric
under the exchange of flavor indices $f\leftrightarrow g$
due to the anti-commuting nature of the fermion fields.
Such condensates can be decomposed as
%\begin{align}
$
	K_{\rm s(ps)}^{fg}(0)
    =
    a_1 t_1^{fg}
    +
    a_3 t_3^{fg}
    +
    a_{13} \omega_{13}^{fg}
    +
    a_{24} \omega_{24}^{fg}
    +
    a_{25} (t_2t_5)^{fg}
    +
    a_{45} (t_4t_5)^{fg}
%\end{align}
$,
where we have introduced
the six independent anti-symmetric matrices 
$t_1$, $t_3$, $\omega_{13}$, $\omega_{24}$,
$t_2t_5$, and $t_4t_5$,
with 
$t_{\mu}={}^{\rm t}\gamma_{\mu}$,
$\omega_{\mu\nu}
=(i/2)[t_{\mu},t_{\nu}]$, 
and $t_5=t_1 t_2 t_3 t_4$.
We have used the representation for the Dirac gamma matrices given by
\begin{align*}
	\gamma_{i}
	=
	\begin{pmatrix}
	0 & i\sigma^{i}
	\\
	-i\sigma^{i} & 0
	\end{pmatrix} \ ,  \quad
	\gamma_{4}
	=
	\begin{pmatrix}
	I_2 & 0
	\\
	0 & -I_2
	\end{pmatrix}  \  ,
\end{align*}
where $\sigma^i$ represent the Pauli matrices
and $I_2$ represents the $2\times 2$ identity matrix.

We calculate the coefficients $a_{1,3,13,24,25,45}$
numerically from the obtained eigenvector and find
%$|a_{13(24)}|/a_{\rm sum}> 99.7\%$
$|a_{13}|/a_{\rm sum}> 0.997$ in $K_{\rm s}^{fg}(0)$ and
$|a_{24}|/a_{\rm sum}> 0.997$ in $K_{\rm (ps)}^{fg}(0)$, where 
$a_{\rm sum}=\sum_{i=1,3,13,24,25,45}|a_{i}|$.
%% which implies that only $a_{13}$ in $K_{\rm s}^{fg}(0)$
%% and $a_{24}$ in $K_{\rm ps}^{fg}(0)$
%% have large absolute values compared to other ones.
This result suggests the following structures
%Figure \ref{fig:adep_mq0} shows
%the flavor dependences of $K_{\rm s(ps)}^{fg}(0)$
%with $\ma=0$ and the second-peak position
%$\mua=\sinh^{-1}(\sin(2\pi/L_{\rm s}))$
%under the standard representation for the Dirac gamma matrices.\footnote{In this
%paper, with $\sigma^i$ and $I_2$ being
%the Pauli matrices and the $2\times 2$ identity matrix, respectively, they are defined by
%\begin{align*}
%	\gamma_{i}
%	=
%	\begin{pmatrix}
%	0 & i\sigma^{i}
%	\\
%	-i\sigma^{i} & 0
%	\end{pmatrix},\quad
%	\gamma_{4}
%	=
%	\begin{pmatrix}
%	I_2 & 0
%	\\
%	0 & -I_2
%	\end{pmatrix}.
%\end{align*}
%}
%This figure clearly shows that the scalar and pseudo scalar
%condensates respectively have the following structures:
\begin{align}
	\label{eq:Ks_flav}
	K^{fg}_{\mathrm{s}}(n)
	=&\,
	\omega_{13}^{fg}
	\kappa_{\rm s}(n) \ ,
	\\
	\label{eq:Kps_flav}
	K^{fg}_{\mathrm{ps}}(n)
	=&\,
	\omega_{24}^{fg}
	\kappa_{\rm ps}(n) \ ,
\end{align}
with
%we have introduced
%$\omega_{\mu\nu}^{fg}
%=(i/2)[t_{\mu},t_{\nu}]^{fg}$, $t_{\mu}={}^{\rm t}\gamma_{\mu}$,
%and 
the flavor-independent coefficients $\kappa_{\rm s(ps)}(n)$,
which give the spatial structure.
%As one can see from these results,
%the condensates are antisymmetric with respect to the exchange of the flavor indices,
%which is a natural consequence 
%of the anti-commutation relation of fermions
%for color-antisymmetric Lorentz-scalar (spin-0)
%condensates.

%% The absence of the patterns other than
%% Eqs.~\eqref{eq:Ks_flav} and \eqref{eq:Kps_flav}
%% is a lattice artifact
Let us discuss the implications of the flavor structure \eqref{eq:Ks_flav} and \eqref{eq:Kps_flav}.
Note first that
the chiral $\mathrm{SU}_{\rm L}(4)\times \mathrm{SU}_{\rm R}(4)$ symmetry
of the continuum theory is reduced on the lattice
to the chiral ${\rm U}(1)$ symmetry $\psi(n)\to e^{i\theta\gamma_5 \otimes t_5}\psi(n)$
in the case of the massless staggered fermions.
%the condensate should appear equally for any choice of the flavors due to
Whether this symmetry is spontaneously broken or not 
depends on the flavor structure of the condensate.
%The action of the massless
%staggered fermion has the symmetry under the chiral ${\rm U}(1)$ transformation:
%$\psi(N)\to e^{i\theta\gamma_5 \otimes t_5}\psi(N)$.
Under the infinitesimal chiral ${\rm U}(1)$ transformation,
Eq.~\eqref{eq:ksps} becomes
\begin{align}
	\label{eq:Kchange}
	K^{fg}_{\rm s(ps)}(n)
	\to
	K^{fg}_{\rm s(ps)}(n)
	+
	i\theta
	\left(
	\sum_{f'}
	t_{5}^{ff'}
	K^{f'g}_{\rm ps(s)}(n)
	+
	\sum_{g'}
	K^{fg'}_{\rm ps(s)}(n)
	t_{5}^{gg'}
	\right) \ .
\end{align}
When $K^{fg}_{\mathrm{s}}(n)$ and $K^{fg}_{\mathrm{ps}}(n)$
have the structure \eqref{eq:Ks_flav}, \eqref{eq:Kps_flav},
we therefore obtain
\begin{align}
	K_{\mathrm{s(ps)}}^{fg}(n)
	\to
	K_{\mathrm{s(ps)}}^{fg}(n)
	+
	2i\theta \omega_{13(24)}^{fg} \kappa_{\rm ps(s)}(n)
\end{align}
%$
%	K_{\mathrm{s(ps)}}^{fg}(N)
%	\to
%	K_{\mathrm{s(ps)}}^{fg}(N)
%	+
%	2i\theta \omega_{13(24)}^{fg} \kappa_{\rm ps(s)}(N)
%$
by using $\omega_{13}t_5=\omega_{24}$
and $\omega_{24}t_5=\omega_{13}$,
%For finite $\theta$,
%this transformation is summarized in terms of $\kappa_{\rm s(ps)}(N)$ as
%\begin{align}
%	\begin{pmatrix}
%		\kappa_{\rm s}(N)
%	\\		
%		\kappa_{\rm ps}(N)
%	\end{pmatrix}
%	\to
%	e^{2i\theta \sigma^{1}}
%	\begin{pmatrix}
%		\kappa_{\rm s}(N)
%	\\		
%		\kappa_{\rm ps}(N)
%	\end{pmatrix}.
%\end{align}
which implies that the chiral $\mathrm{U}(1)$ symmetry is broken spontaneously.
%by the condensates given by Eqs.~\eqref{eq:Ks_flav} and \eqref{eq:Kps_flav}.
This is in contrast to
the other patterns 
$K_{\mathrm{s(ps)}}(n)\sim t_{1},t_3, t_2 t_5, t_4 t_5$
since they are invariant under the chiral $\mathrm{U}(1)$ transformation
as one can show by using Eq.~\eqref{eq:Kchange}.
Note that there is another possibility for condensate
$K^{fg}_{\mathrm{s}}(n)
=\omega_{24}^{fg}\kappa_{\rm s}(n)$
and
$K^{fg}_{\mathrm{ps}}(n)=\omega_{13}^{fg}
\kappa_{\rm ps}(n)$
that breaks the chiral $\mathrm{U}(1)$ symmetry spontaneously.
Thus, our method is capable of
determining not only the symmetry breaking pattern
but also the actual structure of the Cooper pairs
thanks to the fact that we do not have to impose any ansatz on it.
%% not only the discussion about the symmetry breaking
%% but also quantitative analyses of the gap equation
%% without any ansatz for the structure of the Cooper pairs 
%% such as ours are required to determine the structure.
%%
%Thus the symmetry breaking pattern
%does not uniquely determine 
%the condensate structure,
%and quantitative analyses of the gap equation
%without any ansatz for the structure of the Cooper pairs 
%such as ours are required to determine the structure.

\begin{figure}[!t]
  \begin{center}
  	\includegraphics[width=0.8\columnwidth]{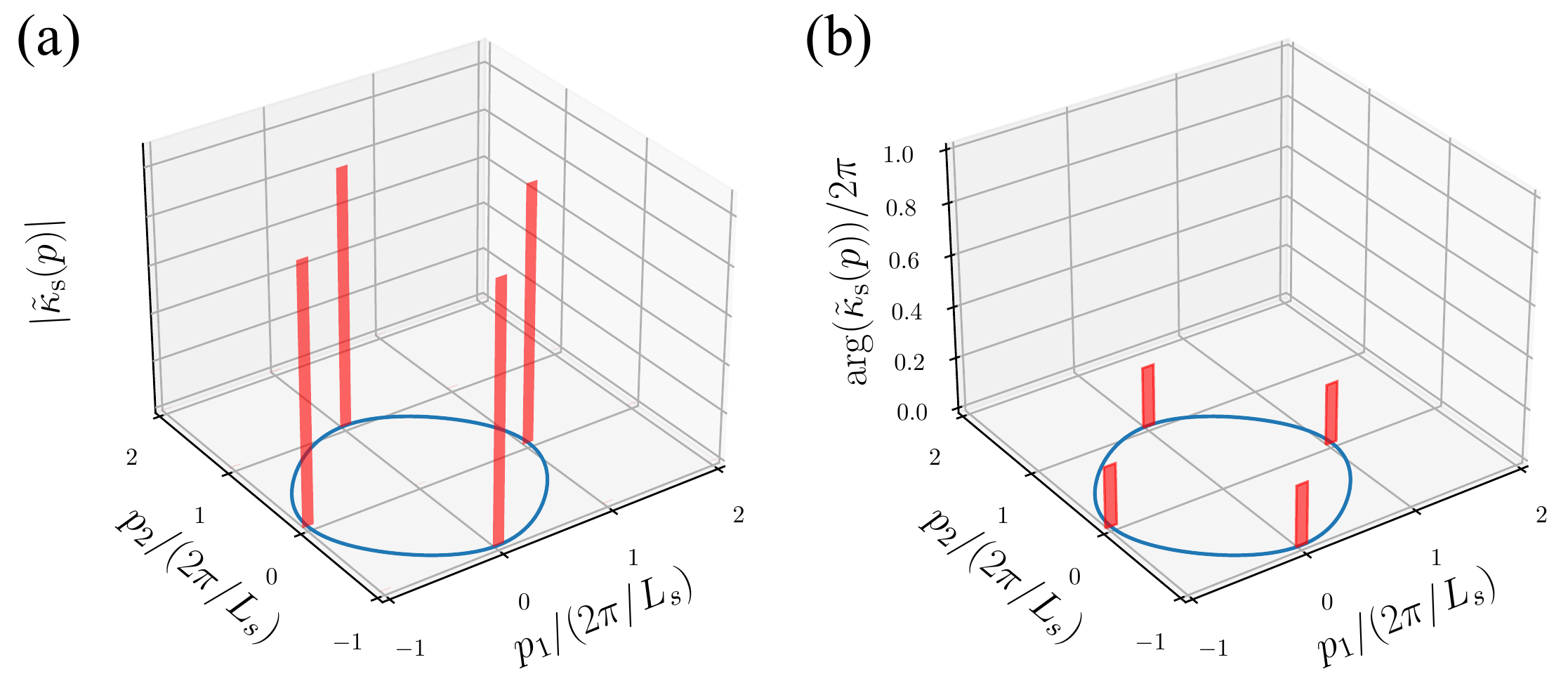}
    \caption{The $p_{1}$ and $p_{2}$ dependence of $\tilde{\kappa}_{\rm s}(p)$
    is shown for $p_{3}=0$ and $p_{4}=\pi/L_{\rm t}$
    with $\ma=0$ and 
    $\mua=\sinh^{-1}(\sin(2\pi/L_{\rm s}))$.
    Figure (a) shows the absolute value 
    $|\tilde{\kappa}_{\rm s}(p)|$ in arbitrary units,
    whereas Fig.~(b)  shows the phase $\arg(\tilde{\kappa}_{\rm s}(p))$
    for the momentum $p$ at which $|\tilde{\kappa}_{\rm s}(p)|$ has large values.
    The blue line represents the Fermi surface for $p_{3}=0$.}
    \label{fig:kappa}
  \end{center}
\end{figure}
We also investigate the spatial structure of the Cooper pairs
represented by $\kappa_{\rm s(ps)}(n)$.
For that, it is convenient to consider the momentum representations
$\tilde{\kappa}_{\rm s(ps)}(p)$.
%are convenient to interpret the results.
Figure \ref{fig:kappa} shows the dependence of
$\tilde{\kappa}_{\rm s}(p)$
on the spatial momenta $p_1$ and $p_2$
for $p_3=0$ and the lowest Matsubara frequency 
$p_4=\pi/L_{\rm t}$
chosen at the position of the second peak.
One can see that $|\tilde{\kappa}_{\rm s}(p)|$
has relatively large values on the Fermi surface,
which is represented by the blue line,
and the values on the Fermi surface are almost the same.
This clearly shows that 
the quarks on the Fermi surface
form the Cooper pairs
and they are spatially isotropic s-waves.
Figure \ref{fig:kappa} (b) shows that
all the modes on the Fermi surface
have the same phase.
This is consistent with the BCS theory 
of superconductivity, where
the phase of the wave function is spatially aligned implying that
the $\mathrm{U}(1)$ particle-number symmetry
is spontaneously broken.
%% Therefore, our result of $\tilde{\kappa}_{\rm s}(p)$
%% follows basic properties of superconductivity.
The results for $\tilde{\kappa}_{\rm ps}(p)$ are qualitatively the same
as those for $\tilde{\kappa}_{\rm s}(p)$.

\section{Critical coupling for the Wilson fermions\label{sec:bc_wil}}
\begin{figure}[!t]
  \begin{center}
  	\includegraphics[width=0.5\columnwidth]{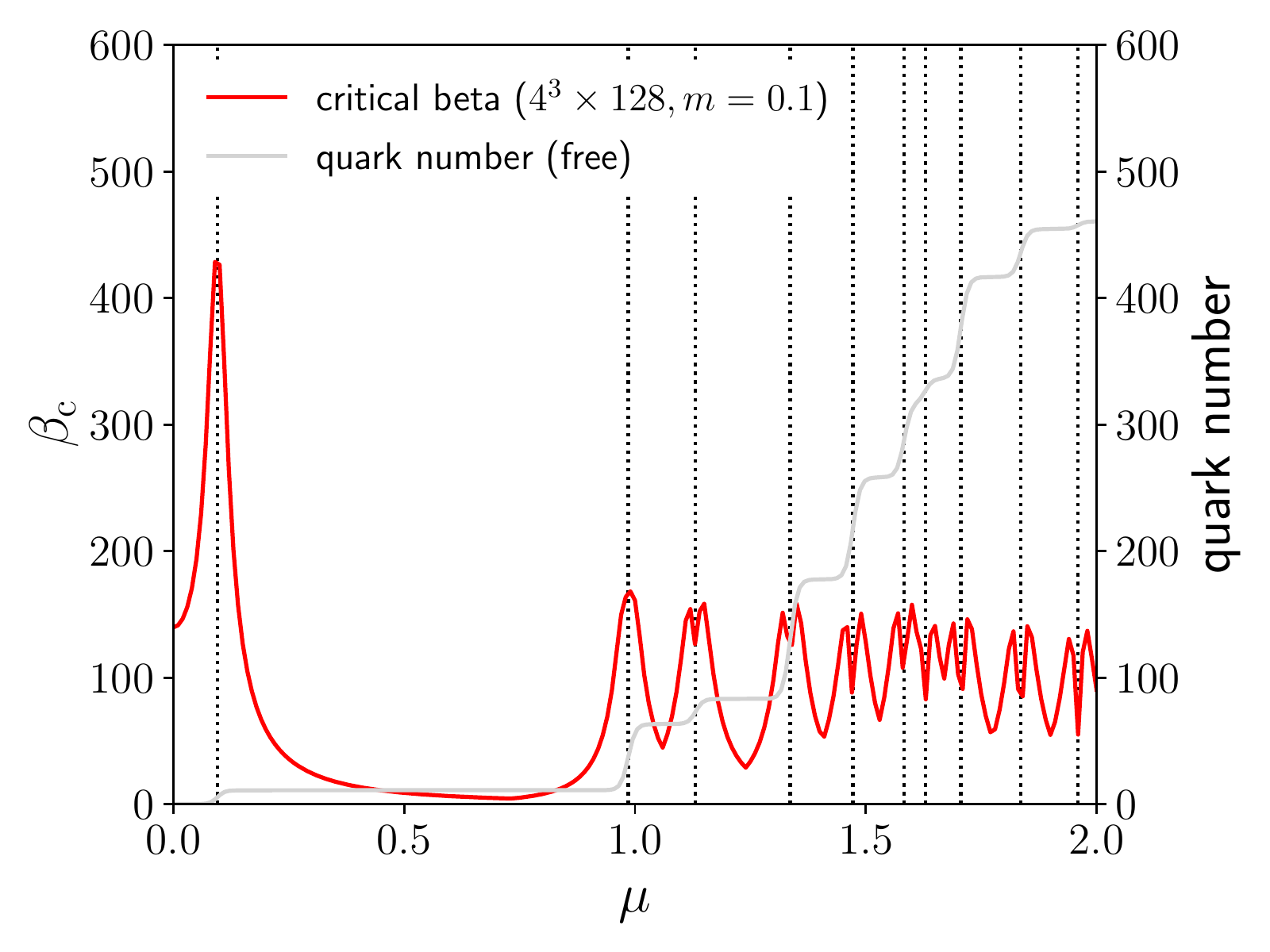}
    \caption{The critical coupling $\beta_{\rm c}$ is plotted as a function of $\mua$
    for the Wilson fermions on a $4^3\times 128$ lattice.
    The gray solid line represents the quark number in the free case,
    whereas the vertical black dotted lines represent
    the position of the energy levels of free quarks.}
    \label{fig:l8t128_wil}
  \end{center}
\end{figure}
It is expected that various types of CSC
appear depending on the number of flavors.
An advantage of the Wilson fermions to the
staggered fermions is that one can choose the number of flavors freely,
although the drawback is that one loses explicit chiral symmetry.
%We present our calculation of 
%in the Wilson-fermion case.
%\sout{We also apply our formalism to the case of
%the Wilson fermions and calculate $\beta_{\rm c}$.}
Figure \ref{fig:l8t128_wil}
shows the result for the critical coupling $\beta_{\rm c}$
with the lattice size
$4^3\times 128$ and $\ma=0.1$.
%corresponding to the hopping parameter $\kappa=0.12195$.
The result does not depend on the number of flavors except for the one-flavor case,
which should be treated in a different manner
%requires some modification of our calculation
since the anti-symmetric flavor structure is not possible.
We also show the particle number of quarks $N_{\rm q}$ in the free case
by the gray solid line and the position of the quark energy levels
%\begin{align}
%	E({\bf p})
%	=
%	2\sinh^{-1}
%	\sqrt{
%	\frac{\sum_{i}\sin^2 p_i +
%	M({\bf p})^2
%	}{4
%	\left(
%	1+
%	M({\bf p})
%	\right)}
%	},
%\end{align}
%with $M({\bf p})=\ma+2\sum_i\sin^2(p_i/2)$,
by the black vertical lines.
As in the case of the staggered fermions,
we can see peaks in $\beta_{\rm c}$
corresponding to the jumps of $N_{\rm q}$ at the energy levels.
On the other hand, we observe more structures
than in the case of staggered fermions,
such as splitting of the peaks at the energy levels higher than the third one.
We are
%In future work, we will
currently investigating the structure of the Cooper pairs
%at these peaks
to clarify the relation to the splitting of the peaks.

\section{Conclusion\label{sec:conc}}
We have investigated color superconductivity (CSC) on the lattice
based on perturbative calculations without any ansatz on the structure of the Cooper pairs.
In our method, the critical coupling $\beta_{\rm c}$
is obtained by calculating the largest eigenvalue of the matrix
that appears in the linearized gap equation,
which is feasible on a finite lattice.
By applying this method to the staggered fermions,
we have obtained $\beta_{\rm c}$
as a function of the chemical potential $\mua$,
which gives the parameter region for CSC.
The result shows that $\beta_{\rm c}$ has peaks at $\mua$
corresponding to the quark energy levels.
We have also investigated the structure of the Cooper pairs at the critical point
from the eigenvector corresponding to $\beta_{\rm c}$ in the massless case.
It turns out that the flavor structure of the (pseudo) scalar condensate thus specified 
breaks the chiral $\mathrm{U}(1)$ symmetry of the staggered fermions spontaneously.
We have also obtained the results of $\beta_{\rm c}$ in the Wilson-fermion case,
which show some splitting of the peaks in contrast to the staggered fermions.

%For completeness of our study,
%Below we list some directions for future investigations.
In the staggered-fermion case,
we have also investigated
other types of condensates such as a pseudo-vector and a tensor
as well as the effect of the quark mass on the Cooper pairs
and the degeneracy of the largest eigenvalue,
which shall be reported in the forth-coming paper.
We hope that our prediction on the parameter region for CSC is useful
in exploring the QCD phase diagram based on first-principles calculations.
We also expect that the structure of the condensate
provides useful information for the construction of 
the order parameter to detect the CSC.
We are currently
trying to observe the CSC on the lattice
%Our another ongoing project is the investigation
by using the complex Langevin method (CLM).
In order to explore the parameter regions suggested in the present work,
we need to extend our previous study of dense QCD
with the staggered fermions \cite{Ito:2020mys} 
to the case with small aspect ratios of the lattice size.
The results for some candidate of the order parameter for the CSC
are presented in Ref.~\cite{Tsutsui:2021pos}.
As for the Wilson fermion,
some basic properties of the CLM 
for $N_{\rm f}=2, 2+1, 3, 4$
on lattices with small aspect ratio
are presented in Ref.~\cite{Namekawa:2021pos}.

\acknowledgments
T.\ Y.\ was supported by the RIKEN Special Postdoctoral Researchers Program.
Y.\ N.\ was supported by JSPS KAKENHI Grant Number JP21K03553.
J.\ N.\ was supported in part by JSPS KAKENHI Grant Number JP16H03988.
S.\ T.\ was supported by the RIKEN Special Postdoctoral Researchers Program.
Numerical computation was carried out on
the Oakbridge-CX provided
by the Information Technology Center at the University of Tokyo
through the HPCI System Research project (Project ID: hp200079, hp210078)
and the Yukawa Institute Computer Facility.

% \bibliographystyle{h-physrev5}
% %\bibliographystyle{JHEP}
% \bibliography{pos_lattice2021_ref}

%\begin{thebibliography}{99}
%\bibitem{...}
%....
%
%\end{thebibliography}

\end{document}